\documentclass[12pt]{article}
\usepackage{pic04}
\usepackage{hyperref}
\usepackage{url}
\usepackage{graphicx}

\begin{document}

\title{\bf BEAUTY PHYSICS WITH $B_s^0$ AND $\Lambda_b^0$}
\author{Rudolf Oldeman \\
for the CDF and \mbox{D\O} collaborations\\
{\em Universit\`a di Roma ``La Sapienza'' and INFN}}
\maketitle

%
%
%
%
%
%
\vspace{4.5cm}
%

\baselineskip=14.5pt

\begin{abstract}
Although the $B_s^0$ and $\Lambda_b^0$ hadrons differ only in the
spectator quarks from the well-studied $B_d^0$ and $B_u^+$ mesons,
they provide a unique window on the physics of the $b$-quark.
With no experiments presently running at the $Z$ pole, hadron colliders
now provide the best opportunity to study the $B_s^0$ and the $\Lambda_b^0$.
The collider experiments at the Tevatron, CDF and \mbox{D\O},
have collected large numbers of $B_s^0$ and $\Lambda_b^0$.
Some of their latest preliminary measurements are presented here, 
including masses, lifetimes, and charmless decays.
Progress is made toward measuring the lifetime difference $\Delta\Gamma_s$
between the $B_s^0$ mass eigenstates
and the oscillation frequency $\Delta m_s$.
\end{abstract}
\newpage

\baselineskip=17pt

\section{Introduction}

Apart from the familiar $B_u^+$ and $B_d^0$ mesons,
the existence of three more weakly decaying $B$ hadrons has been firmly established:
the $B_s^0$ and $B_c^+$ mesons and one baryon, the $\Lambda_b^0$.
Other weakly decaying hadrons have been predicted but have not yet been 
unambiguously observed: the $\Xi_b^0$, the $\Xi_b^-$ and the $\Omega_b^-$.

While our primary physics interest is focused on the $b$-quark, 
there are three good reasons why the spectator quark plays a 
major role in the study of the $b$-quark:
\begin{itemize}
\item 
{\bf The spectator quark can make or break a CP eigenstate}.
A $B_s^0$ has a large branching ratio to the CP-even $D_s^+D_s^-$ final state,
through the Cabbibo-favored \mbox{$b\to c \bar{c} s$} transition.
Since this final state is accessible by both $B_s^0$ and $\bar{B}_s^0$,
it contributes to the lifetime difference between the
heavy and the light $B_s^0$ mass eigenstates.
The equivalent decay of the $B_d^0$ results in a $D_s^+D^-$ final state,
which is flavor-specific, and the lifetime difference in the $B_d^0$ system is negligible.
\item
{\bf The spectator quark can exchange $W$'s with the $b$ quark}. 
The most dramatic consequence of this are oscillations: 
through the exchange of two $W^\pm$, the $B_d^0$ and the $B_s^0$ can
transform into their own anti-particle. 
The $B_s^0$ oscillates more than 20 times faster than the $B_d^0$, whose oscillation 
frequency is suppressed due to the tiny CKM matrix element $V_{td}$.
\item
{\bf The spectator quark can annihilate with the $b$-quark}.
This process is dominated by loop-diagrams involving the top-quark.
The decay of a $B_s^0$ into two muons is expected to occur with a branching
ratio larger by $|V_{ts}|^2/|V_{td}|^2$ compared to $B_d^0\to\mu^+\mu^-$.
\end{itemize}

Rare decays and CP violation are not discussed here, since they are
covered elsewhere in these proceedings~\cite{Olaiya,Kwon}.

\section{Production of  $B_s$ and $\Lambda_b$}
The present $B$ factories operate at the $\Upsilon(4S)$ resonance which
produces only $B_u^+$ and $B_d^0$.
The next resonance, the $\Upsilon(5S)$, is heavy enough to produce 
$B_s^0$ mesons. However, the $B_s^0$ cross-section at the $\Upsilon(5S)$ 
is an order of magnitude smaller than the $B_d^0$ production at the 
$\Upsilon(4S)$~\cite{Lee-Franzini:1990gy},
thus making it challenging to collect a competitive number of $B_s^0$ decays.

There are two other practical means of producing the $B_s$ and the $\Lambda_b^0$:
\begin{itemize}
\item {\bf $e^+e^-$ at the $Z$ pole}. Each of the 4 LEP experiments has recorded
about $880\times10^3$ \mbox{$Z\to b\bar{b}$} events, and have contributed significantly
to our knowledge of the $B_s^0$ and $\Lambda_b^0$.
The \mbox{$Z\to b\bar{b}$} sample produced at the Stanford Linear Collider (SLC) is smaller
by an order of magnitude, but profited from the superior vertex resolution and
from the beam polarization. The latter gives a strong correlation 
between the production hemisphere and the charge sign of the $b$ quark and
provides an efficient flavor-tag, a key advantage for oscillation
studies.
\item {\bf High-energy hadron colliders}.
The Tevatron, a $p\bar{p}$ collider at 1.96\,TeV center of mass energy,
is presently operational and has a 
$b\bar{b}$ cross-section that is about $10^{-3}$ of the total inelastic cross-section.
In 2007 the Large Hadron Collider (LHC) in Gen\`eve will provide an abundant
source of $B$ hadrons. Proton-proton collisions at  a center-of-mass 
energy of 14\,TeV produce a $b\bar{b}$ pair roughly every hundred
interactions.
\end{itemize}
In both of the above mentioned cases, the production fractions
of $B_s^0$ and $\Lambda_b^0$ are approximately 10\% each, while $B_c^+$ production
is suppressed at the $10^{-3}$ level.

\section{Reconstruction of $B$-decays}
The types of $B$ decays that can be reconstructed 
at a hadron collider can be distinguished into three classes:
\begin{itemize}
\item
{\bf Semileptonic decays}, for example \mbox{$B_s^0\to D_s^-\mu^+\nu_\mu$}.
These have large branching fractions,
and large yields can be obtained simply by triggering on a lepton, 
but the missing neutrino prohibits complete reconstruction.
Moreover, semileptonic decays cannot result in a CP eigenstates, 
and do not give access to many of the most interesting $B$ physics channels.
\item
{\bf $B$ decays with a $J/\psi$ in the final state}, for example \mbox{$B_s^0\to J/\psi\phi$}.
These have the advantage of providing a clear signature for the trigger,
through the dimuon decay of the $J/\psi$. 
However, the sum of all branching ratios with a $J/\psi$ in the final state
is only slightly more than one percent.
\item 
{\bf Hadronic decays}, such as \mbox{$B_s^0\to D_s^-\pi^+$}, \mbox{$B_s^0\to K^+K^-$}.
These constitute about three quarters of all $B$ decays and offer a rich variety of $B$ physics.
However, they are difficult to distinguish from the
overwhelming background of hadronic interactions without heavy flavor.
Its main signature are displaced tracks,
and to collect large samples of fully reconstructed hadronic $B$ decays
requires specialized triggers that are capable of reading out and processing
data from a silicon vertex detector at high speed.

\end{itemize}

\section{Present and future experiments}
Both the CDF and the \mbox{D\O} detectors are well equipped for a rich
$B$ physics program: both have silicon detectors, high resolution trackers
in a magnetic field, and lepton identification. 
\mbox{D\O} profits from its hermetic muon coverage and its 
efficient tracking in the forward regions, 
giving a high sensitivity for semileptonic and $J/\psi$ modes.
CDF has better track momentum resolution,
a high-bandwidth silicon track trigger, and
particle identification capabilities through Time-of-Flight counters and
$dE/dx$ measurements in its main drift chamber.

Two new specialized $B$ experiments may dramatically improve our 
knowledge of the $B_s^0$ and the $\Lambda_b^0$:
the LHCb experiment, starting in 2007 at the LHC,
and the BTeV experiment, starting in 2009 at the Tevatron.
Both experiments use a dipole spectrometer and are instrumented
at small angles with respect to the beam.
Using the forward region has many advantages:
\begin{itemize}
\item The useful cross-section is higher, because of the large acceptance at small transverse momentum.
\item Often both $B$'s are in the detector acceptance, giving a high efficiency for flavor-tagging.
\item The boost in the direction of the beam allows to accurately measure the decay time
in the beam direction.
\item The forward detector geometry allows to install Rich Imaging Cerenkov detectors
for superb particle identification.
\end{itemize}

\section{Mass measurements}
\begin{figure}[htb]
\begin{center}
\includegraphics[width=6cm]{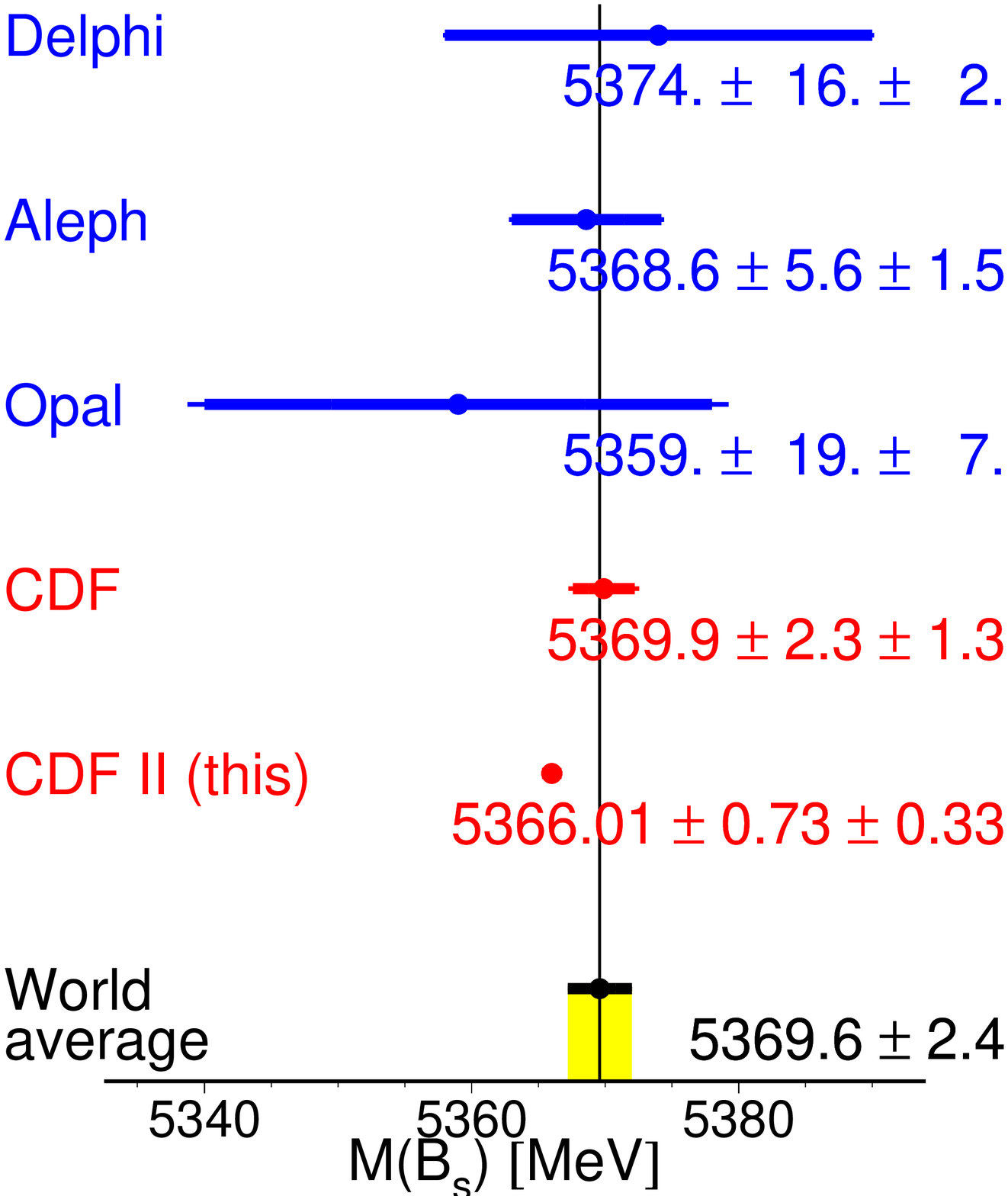}
\includegraphics[width=6cm]{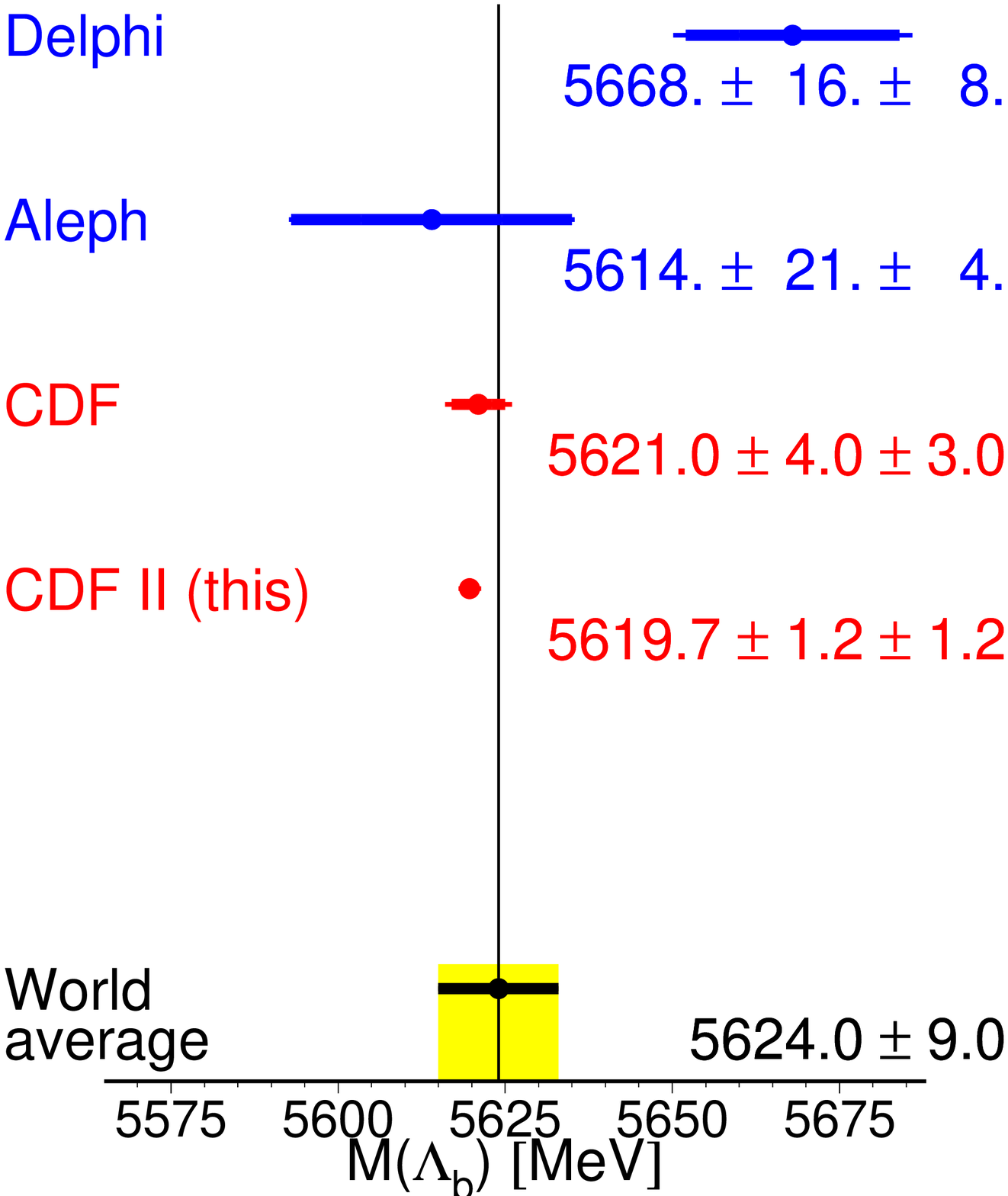}
 \caption{\it
	Preliminary CDF measurements of $B$ hadron masses in the 
        $B_s^0\to J/\psi\phi$ and $\Lambda_b^0\to J/\psi\Lambda$ channels.
    \label{fig:mass} }
\end{center}
\end{figure}

A typical 'mass peak' of 200 fully reconstructed events with an experimental resolution of 15\,MeV gives a mass measurement
with a statistical uncertainty of $\approx$1\,MeV.
Both $B_s^0$ and $\Lambda_b^0$ have now been observed in fully reconstructed decay modes, 
but the $B_c^+$ has only been observed in the semileptonic decay \mbox{$B_c^+\to J/\psi\ell^+\nu_\ell$},
and the mass has an uncertainty of more than 400\,MeV. 
This can be dramatically improved by observing the $B_c^+$ in fully reconstructed
decay modes such as \mbox{$B_c^+\to J/\psi \pi^+$}.
Preliminary $B_s^0$ and $\Lambda_b^0$ mass measurements from CDF, shown in Figure~\ref{fig:mass},
significantly improve the previous best measurements.
To achieve mass measurements at the 1\,MeV level, the mass scale needs to be understood to
$10^{-4}$.
This has been achieved by calibrating on \mbox{$J/\psi\to\mu^+\mu^-$} decays,
which are copiously produced in hadron colliders.

\section{Lifetime measurements}

To first order, $B$-hadron lifetimes are determined by the fastest decaying quark:
\begin{equation}
\tau(B_u^+)\approx\tau(B_d^0)\approx\tau(B_s^0)\approx\tau(\Lambda_b^0)\gg\tau(B_c^+)
\end{equation}
Spectator effects can be calculated in the Heavy Quark Expansion, but have been determined to be small: 
the dominant contributions scale with $(\Lambda_{QCD}/m_b)^3$. 
A recent calculation includes $m_b^{-4}$ contributions~\cite{Gabbiani:2004tp} and finds:
\begin{equation}
\frac{\tau(B_u^+)}{\tau(B_d^0)}=1.09\pm0.03, \,\,\,  
\frac{\tau(B_s^0)}{\tau(B_d^0)}=1.00\pm0.01, \,\,\,
\frac{\tau(\Lambda_b^0)}{\tau(B_d^0)}=0.87\pm0.05.
\end{equation}
The best lifetime measurements of the $B_s^0$ and the $\Lambda_b^0$ come from semileptonic decays 
at CDF-I and LEP. 
The current World Average~\cite{PDG} is:
\begin{equation}
\tau(B_s^0)=1.46\pm0.06\,{\rm ps} \,\,\,{\rm and} \,\,\,
\tau(\Lambda_b^0)=1.23\pm0.08\,{\rm ps}.
\end{equation}
Semileptonic measurements, however, suffer from incomplete reconstruction due to the
missing neutrino. This introduces irreducible systematic uncertainties both from the 
production model and from the decay model.
Fully reconstructed $B$ decays are not affected by model-dependencies, since the
lifetime is measured on an event-by-event basis, but they provide smaller statistics.
Both \mbox{D\O} and CDF have recently measured the lifetimes of the $B_s^0$ and $\Lambda_b^0$ 
in fully reconstructed modes with a precision similar to the semileptonic measurements:
\begin{eqnarray}
{\rm CDF}\, 220\,{\rm pb}^{-1}: \tau(B_s^0\to J/\psi\phi)          & = & 1.37\pm0.10\pm0.01\,{\rm ps}, \\
{\rm \mbox{D\O}} \, 115\,{\rm pb}^{-1}: \tau(B_s^0\to J/\psi\phi)  & = & 1.19\pm0.19\pm0.14\,{\rm ps}, \\
{\rm CDF}\,  65\,{\rm pb}^{-1}: \tau(\Lambda_b^0\to J/\psi\Lambda) & = & 1.25\pm0.26\pm0.10\,{\rm ps}.
\end{eqnarray}
The last one represents the first measurement of the $\Lambda_b^0$ lifetime from fully reconstructed decays.
In the near future these measurements will be updated with more data, 
and we can expect lifetime measurements from semileptonic and hadronic modes.

\section{The $B_s^0$ lifetime difference}

Because of mixing, the time evolution of neutral $B$ mesons is not governed
by the flavor eigenstates $B^0$, $\bar{B}^0$, but by the mass eigenstates $B_L$, $B_H$,
which may differ not only in mass, but also in decay width by $\Delta\Gamma=\Gamma_L-\Gamma_H$.
For non-zero $\Delta\Gamma$, the decay time distribution follows 
a double instead of a single exponential.
If the $B_s$ mixing phase is as small as predicted, $\phi_s\approx 0.03$,
the $B_s^0$ mass eigenstates coincide almost exactly with the CP eigenstates.
A significant lifetime difference is then expected from
the Cabibbo-favored \mbox{$b\to c\bar{c}s$} transition 
that results in a large fraction of final states that are
CP even and common to $B_s^0$ and $\bar{B}_s^0$.

A recent calculation~\cite{Ciuchini:2003ww} predicts $\Delta\Gamma_s/\Gamma_s=0.074\pm0.024$,
consistent with the experimental world-average value 
$\Delta\Gamma_s/\Gamma_s=0.07^{+0.09}_{-0.07}$,
obtained under the assumption that $\tau(B_s^0)=\tau(B_d^0)$,
or $\Delta\Gamma_s/\Gamma_s=0.16^{+0.15}_{-0.16}$ without this constraint~\cite{HFAG}.

Three methods are available to measure $\Delta\Gamma_s$:
\begin{enumerate}
\item
Take a CP-mixed decay, and fit the lifetime distribution to a double exponential.
A disadvantage is that this is sensitive to $(\Delta\Gamma_s)^2$, 
making it difficult to probe small values of  $\Delta\Gamma_s$.
\item
Compare the $B_s^0$ lifetime in a CP-even to a CP-odd or CP-mixed state. 
$B$ decays to two spin-1 particles, such as \mbox{$B_s^0\to J/\psi\phi$},
can be decomposed through an angular analysis into CP-odd and CP-even states.
\item
Since the lifetime difference is dominated by the decay \mbox{$B_s^0\to D_s^{(*)+}D_s^{(*)-}$},
measurements of these branching fractions provide an indirect measurement of $\Delta\Gamma_s$.
\end{enumerate}

CDF has recently completed a preliminary angular analysis of \mbox{$B_s^0\to J/\psi\phi$}, 
extracting three complex amplitudes, the short-lived CP-even $A_0$ and $A_{||}$,
and the long-lived CP-odd $A_T$, finding:
\begin{eqnarray}
A_0    &=& 0.767\pm0.045\pm0017, \\
A_{||} &=& (0.424\pm0.118\pm0.013)e^{(2.11\pm0.55\pm0.29)i}, \\
A_T    &=& 0.482\pm0.104\pm0.014.
\end{eqnarray}
This shows that \mbox{$B_s^0\to J/\psi\phi$} is mostly CP-even, 
but also has a significant CP-odd component, making it possible to measure $\Delta\Gamma_s$ from this channel alone.

\section{Charmless decays}

\begin{figure}[htb]
\begin{center}
\includegraphics[width=10cm,clip]{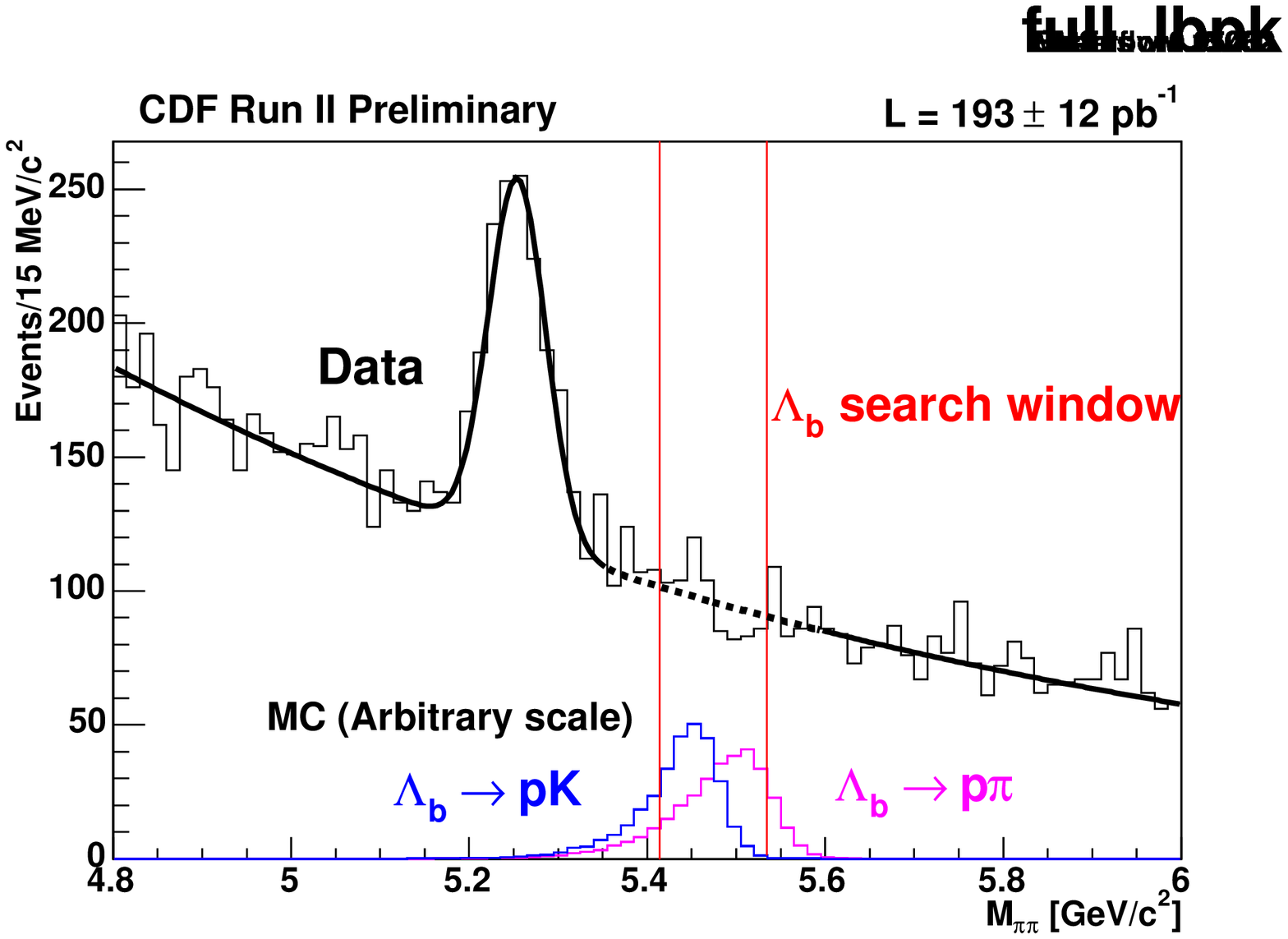}
 \caption{\it
     $B\to h^+h^-$ plot with the pion mass assumption for both tracks, indicating the search window for \mbox{$\Lambda_b^0\to pK^-,p\pi^-$}.
	The peak at $m_{\pi\pi}\approx5.27$\,GeV is dominated by $B_d^0\to K^+\pi^-$ decays used for normalization.
    \label{fig:mpipi_lambdab} }
\end{center}
\end{figure}

The strongly suppressed \mbox{$b\to u$} transitions probe the CKM matrix element $V_{ub}$ and its phase, 
often called $\gamma$.
In practice, interfering contributions from penguin decays complicate
precision measurements of $\gamma$ from charmless $B$ decays.
Comparing various two-body decays of $B_s^0$ and $B_d^0$ allows to disentangle 
the tree and penguin contributions~\cite{Fleischer:1999nz}.
The challenge of reconstructing 2-body $B$ decays at a hadron collider resides both in
rejecting large backgrounds and in distinguishing
for example a \mbox{$B_d^0\to\pi^+\pi^-$} decay from a \mbox{$B_s^0\to K^+K^-$} decay 
without strong particle identification.
Using specific ionization in their drift chamber, CDF achieves a $\pi/K$ separation of $1.15\,\sigma$,
enough to disentangle the four main contributions to the peak in their $m(\pi^+\pi^-)$ histogram,
and measures $Br(B_s^0\to K^+K^-)/Br(B_d^0\to K^+\pi^-)=2.71\pm0.73\pm0.88$.
This measurement  is related to the direct CP violation in
\mbox{$B_d^0\to\pi^+\pi^-$} decays and was found to agree with the Standard Model expectation~\cite{London:2004uj}.
The same data have also been used to search for the charmless decay \mbox{$\Lambda_b^0\to p\pi^-, pK^-$}.
The predicted branching ratios,
\mbox{$(1.4-1.9)\times10^{-6}$} for \mbox{$\Lambda_b^0\to pK^-$} and
\mbox{$(0.8-1.2)\times10^{-6}$} for \mbox{$\Lambda_b^0\to p\pi^-$}~\cite{Mohanta:2000nk},
are considerably lower than the best experimental upper limit:
$Br(\Lambda_b^0\to p\pi^-, pK^-)\leq 50\times10^{-6}$ at 90\%CL~\cite{Buskulic:1996tx}.
CDF finds no evidence for a signal and improves the upper limit to 
$Br(\Lambda_b^0\to p\pi^-, pK^-)\leq 22\times10^{-6}$ at 90\%CL.
Figure~\ref{fig:mpipi_lambdab} shows the \mbox{$B\to h^+h^-$} meson signal, the $\Lambda_b$ search window 
and simulated \mbox{$\Lambda_b^0\to pK^-,p\pi^-$} signals.
Particularly interesting charmless decays come from the pure penguin \mbox{$b\to s\bar{s}s$} transition, since
the Belle collaboration has observed a $3.5\sigma$ deviation
from the expected value of the weak phase in the $B_d^0\to \phi K^0_S$ channel~\cite{Abe:2003yt}.
CDF observes 12 \mbox{$B_s^0\to\phi\phi$} candidates with an expected background of 
$1.95\pm0.62$ events, constituting the first \mbox{$b\to s\bar{s}s$} in $B_s^0$ decays.
They measure \mbox{$Br(B_s^0\to\phi\phi)=(14\pm6_{stat}\pm2_{syst}\pm5_{Br})\times10^{-6}$},
where the last uncertainty comes from \mbox{$Br(B_s^0\to J/\psi\phi)$} that is used as a normalization mode.
The measured branching fraction is consistent with the wide range of predictions,
which cover the range $(0.4-37)\times10^{-6}$~\cite{Chen:1998dt,Li:2003he}.
The large branching ratio, combined with a distinct and low-background experimental signature,
promises a bright future for this channel 
including angular analyses, measurements of $\Delta\Gamma_s$, and CP violation.

\section{$B_s^0$ oscillations}

\begin{figure}[htb]
\begin{center}
\includegraphics[width=7cm]{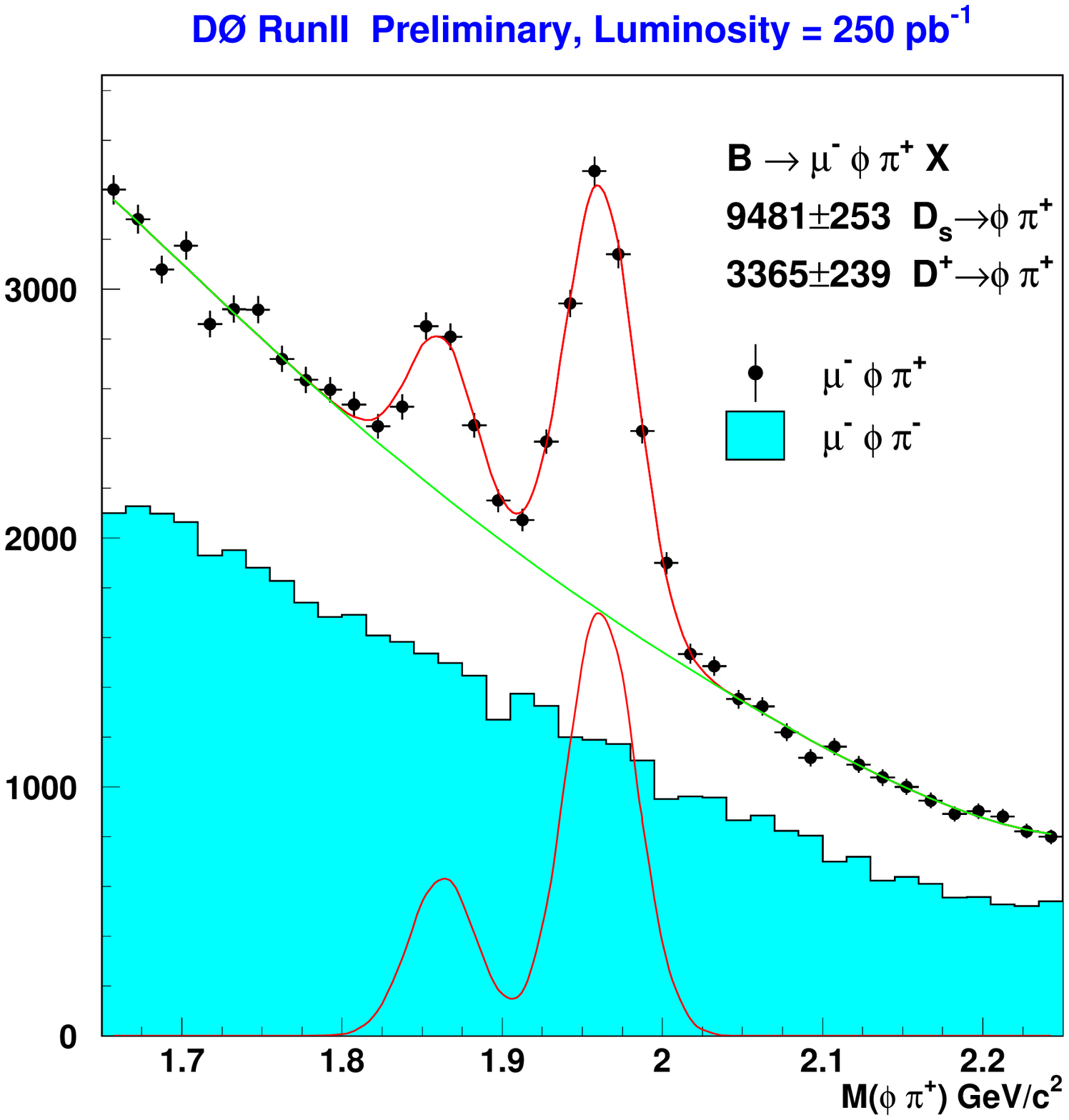}
\includegraphics[width=7cm]{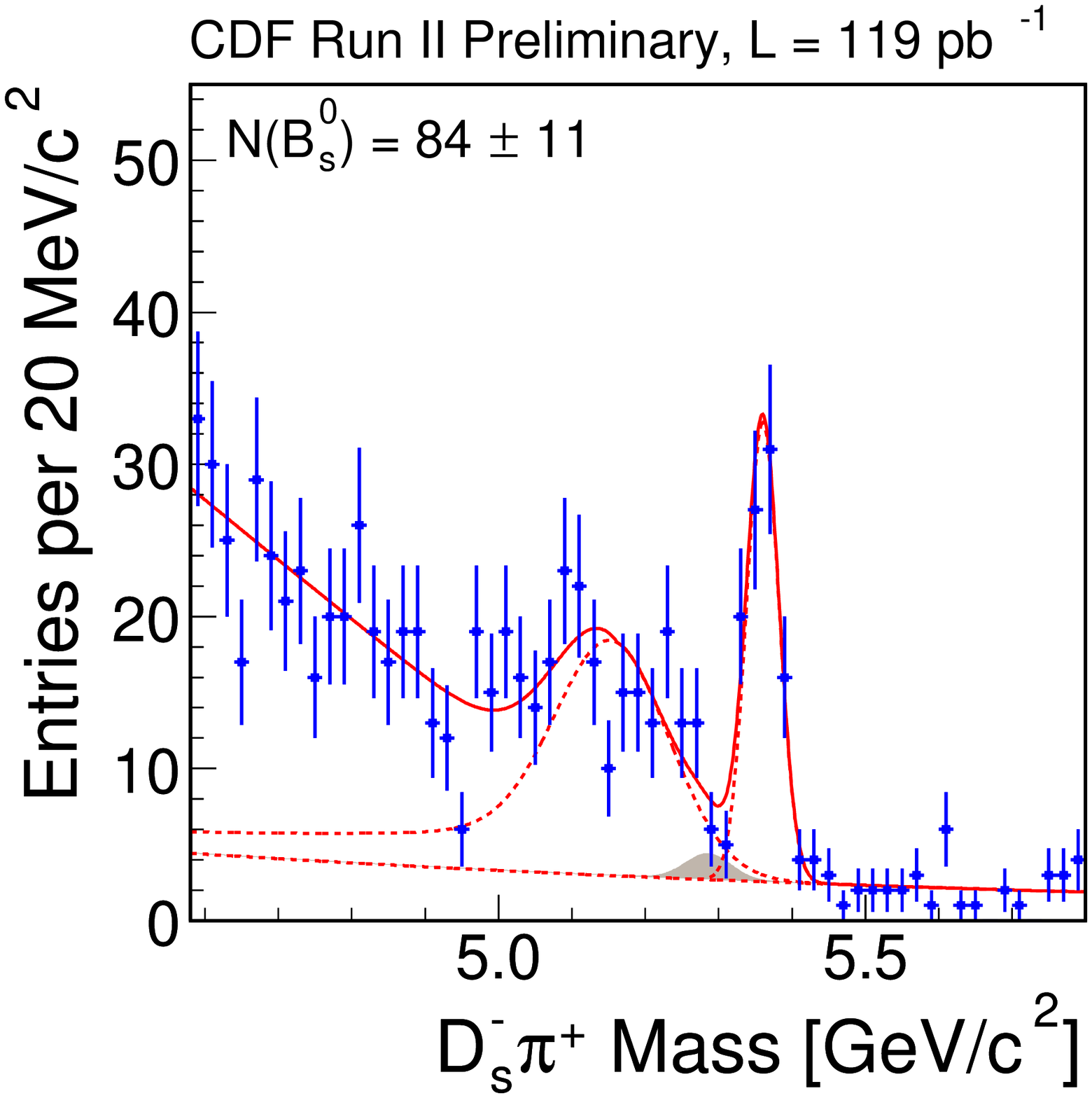}
 \caption{\it
    semileptonic $B_s^0$ yields from \mbox{D\O} (left) and fully hadronic yields from CDF (right).
    \label{fig:Bs_yields} }
\end{center}
\end{figure}

The well-measured $B_d^0$ oscillations provide a measurement of the CKM element $|V_{td}|$, 
but the extraction is plagued by theoretical uncertainties.
A more accurate measurement can be obtained by measuring also the 
$B_s^0$ oscillation frequency and use~\cite{Schneider:2004hc}
\begin{equation}
\frac{\Delta m_s}{\Delta m_d}=\frac{m(B_s^0)}{m(B_d^0)}\left(1.15\pm0.06^{+0.12}_{-0.00}\right)^2\left|\frac{V_{ts}}{V_{td}}\right|^2,
\end{equation}
where the last (asymmetric) uncertainty comes from the chiral extrapolation.
Seen the other way around, the Standard Model gives an accurate prediction of $\Delta m_s$,
and many new physics models allow significantly larger values~\cite{Barger:2004qc,Ball:2003se}.
In addition, a precise measurement of $\Delta m_s$ is a prerequisite for many time-dependent 
CP violation studies in the $B_s^0$ system.
The present experimental lower limit is $\Delta m_s \geq 14.4$\,ps$^{-1}$~\cite{PDG},
implying more than three full $B_s^0$ oscillation cycles within one lifetime.
Because of these very fast oscillations, 
the precise  measurement of the proper decay time is of crucial importance:
since the uncertainty on the oscillation amplitude scales like
$\sigma(A)\propto e^{(\sigma_t\Delta m_s)^2/2}$,
a proper time resolution larger than 67\,fs$^{-1}$ seriously affects the sensitivity 
above 15\,ps$^{-1}$.
The first ingredient for mixing is a large $B_s^0$ yield.
Figure~\ref{fig:Bs_yields} shows the large yields from
\mbox{D\O} in \mbox{$B_s^0\to D_s^-\mu^+\nu_\mu$} decays
and the first \mbox{$B_s^0\to D_s^-\pi^+$} signal from CDF, which is much smaller in statistics,
but provides a more accurate measurement of the proper decay time.

Another crucial ingredient for any mixing measurement is flavor tagging, 
to determine whether a $B_s^0$ or a $\bar{B}_s^0$ was produced.
This information can be obtained either from the fragmentation tracks of the 
$B_s^0$ under study (``same side tag''), or from the decay products of the 
$b$ quark that is produced in association with the $B_s^0$ (``opposite side tag'').
The effectiveness of a flavor tagger is usually expressed in its efficiency $\varepsilon$
and its ``Dilution factor'' $D=1-2W$, where $W$ is the fraction of wrong charge assignments.
The statistical power of a flavor tagger scales as $\varepsilon D^2$.
Contrary to $B$ physics at the $Y(4S)$, where values of $\varepsilon D^2\approx30\%$
are readily achieved, the flavor taggers at hadron colliders rarely exceed an 
$\varepsilon D^2$ of one percent.

Both \mbox{D\O} and CDF have shown non-zero dilutions for opposite side muon and jet-charge taggers.
Both have also shown powerful tagging using fragmentation particles associated with the $B_u^+$.
However, for the $B_s^0$, the flavor information is typically carried by a kaon, 
and it requires good $\pi/K$ separation to use same-side taggers for the $B_s^0$.

CDF and \mbox{D\O} have produced preliminary $B_d^0$ mixing measurements.
\mbox{D\O} measures $\Delta m_d=0.506\pm0.055\pm0.049$\,ps$^{-1}$ using semileptonic
$B_d^0$ decays and an opposite side muon tag.
CDF measures $\Delta m_d=0.55\pm0.10\pm0.01$\,ps$^{-1}$ using fully reconstructed 
hadronic decays and a same-side tag.

\section{Conclusions}

The physics of the $B_s$ and $\Lambda_b$ provide a unique window 
on $B$ physics that is not accessible at the $\Upsilon(4S)$.
New measurements of masses, lifetimes and observations in new decay modes
have recently come available from the collider experiments at the Tevatron.
These, and future measurements of the $B_s^0$ mixing parameters $\Delta m_s$ and $\Gamma_s$
will determine the physics opportunities at the next generation hadron $B$ physics experiments
at hadron colliders.

\end{document}